\documentclass{PoS}
\usepackage{amsmath,amssymb,bbm,color}
\usepackage{graphicx}
\usepackage{epsfig}
\usepackage{euscript}
\usepackage{pslatex}
\usepackage{fancybox}
\usepackage{enumerate}
\usepackage{widetext}
\usepackage{subfigure}
\title{Prediction for the Cosmological Constant
and Constraints on SUSY GUTS: Status Report for Resummed Quantum Gravity}

\ShortTitle{Prediction for the Cosmological Constant}

\author{\speaker{B.F.L. Ward}\thanks{On Research Leave from Baylor University, Waco, TX, USA, Jan. 4 - Jul. 31,2018, at Werner-Heisenberg-Institut, Max-Planck-Institut fuer Physik, Muenchen, Germany }\\
        Baylor University, Waco, TX, USA\\
        E-mail: \email{BFL\_Ward@baylor.edu}}

\abstract{We give a status report on the theory of resummed quantum gravity. We recapitulate the use of  our resummed quantum gravity approach to Einstein's general theory of relativity to estimate the value of the cosmological constant as $\rho_\Lambda=(0.0024{\mathrm{eV}})^4$. The estimate is made  in the context of the Planck scale cosmology formulation of Bonanno and Reuter. We discuss the constraints on susy GUT models that follow from the closeness of the estimate to experiment. Various consistency checks on the calculation are addressed and we use the Heisenberg uncertainty principle to remove a large part of the remaining uncertainty in our estimate of $\rho_\Lambda$.}
\centerline {\bf BU-HEPP-18-08, Nov., 2018}
\FullConference{ICHEP2018, 39th International Conference on High Energy Physics\\
		5-11 July 2018\\
		Seoul, S. Korea}
\begin{document}
\baselineskip=10.8pt
\section{Introduction}
\label{intro}
We start with the basic question which our title engenders, ``What is resummation in quantum field theory?'' For comparison we illustrate ``summation'' with the elementary example:
\begin{equation}
\frac{1}{1-x} = \sum_{n=0}^{\infty}x^n.
\label{eq-elmtry}
\end{equation} 
The geometric series is summed to infinity to yield the analytic result that is well-defined except for a pole at x=1. Indeed, even though the mathematical tests for convergence of the 
series would only guarantee convergence for $|x|$ <1, the result of the summation yields a function that is well-defined in the entire complex plane except for the simple pole at x=1. 
We see that infinite order summation can yield behavior very much improved from what one sees order-by-order in the series.\par
We are thus motivated to resum series that are already being summed to seek improvement in our knowledge of the represented function. We illustrate this as follows:
\begin{equation}
\sum_{n=0}^{\infty}C_n \alpha_s^n \begin{cases}&= F_{\rm RES}(\alpha_s)\sum_{n=0}^{\infty} B_n \alpha_s^n,\; \text{\rm EXACT}\\
                                                                                                                                      &\cong  G_{\rm RES}(\alpha_s)\sum_{n=0}^{\infty} \tilde{B}_n \alpha_s^n,\; \text{\rm APPROX.}\end{cases}
                                                                                                                                      \label{eq-res}
\end{equation}
The original Feynman series for a process under study is on the LHS. Two versions of resumming this original series are on the RHS. One, labeled exact, is an exact re-arrangement of the original series. The other, labeled approx., only agrees with the LHS to some fixed order N in the expansion parameter $\alpha_s$. Which version is to be preferred has been discussed for some time~\cite{frits-ichep88}. A related more general discussion now occurs for quantum gravity.\par
We may ask whether quantum gravity is even calculable in relativistic quantum field theory? There are various answers. String theory~\cite{strgthy} argues the answer is no, the true fundamental theory entails a one-dimensional Planck scale superstring. Loop quantum gravity~\cite{lpqg} also argues the answer is no, the fundamental theory entails a space-time foam with a Planck scale loop structure. The Horava-Lifshitz theory~\cite{horva} also argues the answer is no because the fundamental theory requires Planck scale anisoptropic scaling for space and time. Kreimer~\cite{kreimer} suggests that quantum gravity is leg-renormalizable, so that the answer is yes. Weinberg~\cite{wein1} suggests that quantum gravity may be asymptotically safe, with an S-matrix
that depends only on a finite number of observable parameters, due to
the presence of a non-trivial UV fixed point, with a finite dimensional critical surface; this amounts to an answer of yes. The authors  in Refs.~\cite{reutera,laut,reuterb,reuter3,litim,perc},
using Wilsonian~\cite{kgw} field-space exact renormalization 
group methods, obtain results which support Weinberg's  UV fixed-point. Ref.~\cite{ambj} also gives support to Weinberg's
asymptotic safety suggestion.\par
In what follows, we extend the YFS~\cite{yfs,jad-wrd} version\footnote{YFS-type soft resummation and its extension to quantum gravity was also worked-out by Weinberg in Ref.~\cite{sw-sftgrav}.} of the exact example to resum the Feynman series for the Einstein-Hilbert Lagrangian for quantum gravity. Just as we see in the example in eq.(\ref{eq-elmtry}), the resultant  resummed theory, resummed quantum gravity (RQG), is very much better behaved in the UV compared to what one would estimate from that Feynman series.
We have shown~\cite{bw1,bw2,bw2a,bw2i} that the RQG realization of quantum gravity leads to Weinberg's UV-fixed-point behavior for the dimensionless
gravitational and cosmological constants and that the resummed theory is actually UV finite. RQG and the latter results are reviewed in Section 2.\par
We show further that the RQG theory, taken together with the Planck scale inflationary~\cite{guth,linde} cosmology formulation in Refs.~\cite{reuter1,reuter2}\footnote{The authors in Ref.~\cite{sola1} also proposed the attendant 
choice of the scale $k\sim 1/t$ used in Refs.~\cite{reuter1,reuter2}.} from the 
asymptotic safety approach to quantum gravity in 
Refs.~\cite{reutera,laut,reuterb,reuter3,litim,perc}, allows us to predict~\cite{drkuniv} the cosmological constant $\Lambda$. Due to the prediction's closeness to the observed value~\cite{cosm1,pdg2008}, we discuss its reliability and argue~\cite{eh-consist} that its uncertainty is at the level of a factor of ${\cal O}(10)$. There follow constraints on susy GUT's.
The Planck scale cosmology that we use and the latter results are presented in Section 3.\par
%In Refs.~\cite{ambj}, we note that causal dynamical triangulated lattice methods have been used to show more evidence for Weinberg's asymptotic safety behavior\footnote{At the expense of violating Lorentz invariance, the model in Ref.~\cite{horva} realizes many aspects
%of the effective field theory implied by the anomalous dimension of 2 at the Weinberg
%UV-fixed point.}.
%%Our discussion is presented as follows. Section 2 contains an overview of the theory of resummed quantum gravity. In Section 3, we review the Planck scale cosmology
%%of Refs.~\cite{reuter1,reuter2} and our predictions~\cite{bw2i} for the fixed point values of
%%the dimensionless gravitational and cosmological constants. We use the latter results to predict the cosmological constant $\Lambda$. We address consistency checks on the prediction and present arguments that its error is at the level of ${\cal O}(10)$. Constraints on susy GUT's implied by our prediction for $\Lambda$ are noted.\par
\section{Overview of Resummed Quantum Gravity}
The Standard Theory~\footnote{We follow D.J. Gross~\cite{djg1} and call the Standard Model the Standard Theory henceforth.} of elementary particles contains many point particles. To explore their graviton interactions, we treat spin as an inessential complication~\cite{mlgbgr} and consider the Higgs-gravition extension of the Einstein-Hilbert theory,
already studied in Refs.~\cite{rpf1,rpf2}, as exemplary:
\begin{equation}
\begin{split}
{\cal L}(x) &= -\frac{1}{2\kappa^2} R \sqrt{-g}
            + \frac{1}{2}\left(g^{\mu\nu}\partial_\mu\varphi\partial_\nu\varphi - m_o^2\varphi^2\right)\sqrt{-g}\\
            &= \quad \frac{1}{2}\left\{ h^{\mu\nu,\lambda}\bar h_{\mu\nu,\lambda} - 2\eta^{\mu\mu'}\eta^{\lambda\lambda'}\bar{h}_{\mu_\lambda,\lambda'}\eta^{\sigma\sigma'}\bar{h}_{\mu'\sigma,\sigma'} \right\}\\
            & + \frac{1}{2}\left\{\varphi_{,\mu}\varphi^{,\mu}-m_o^2\varphi^2 \right\} -\kappa {h}^{\mu\nu}\left[\overline{\varphi_{,\mu}\varphi_{,\nu}}+\frac{1}{2}m_o^2\varphi^2\eta_{\mu\nu}\right]\\
            &  - \kappa^2 \left[ \frac{1}{2}h_{\lambda\rho}\bar{h}^{\rho\lambda}\left( \varphi_{,\mu}\varphi^{,\mu} - m_o^2\varphi^2 \right) - 2\eta_{\rho\rho'}h^{\mu\rho}\bar{h}^{\rho'\nu}\varphi_{,\mu}\varphi_{,\nu}\right] + \cdots\;.\\
\end{split}
\label{eq1-1}
\end{equation}
$R$ is the curvature scalar,  $g$ is the determinant of the metric
of space-time $g_{\mu\nu}\equiv\eta_{\mu\nu}+2\kappa h_{\mu\nu}(x) $, and $\kappa=\sqrt{8\pi G_N}$.
We expand~\cite{rpf1,rpf2} about Minkowski space
with {\small$\eta_{\mu\nu}=\text{diag}\{1,-1,-1,-1\}$}.
$\varphi(x)$ is the physical Higgs field as
our representative scalar field for matter and
$\varphi(x)_{,\mu}\equiv \partial_\mu\varphi(x)$.
%and $g_{\mu\nu}(x)=\eta_{\mu\nu}+2\kappa h_{\mu\nu}(x)$
%where we follow Feynman and expand about Minkowski space
%so that $\eta_{\mu\nu}=\text{diag}\{1,-1,-1,-1\}$.
We have introduced Feynman's notation
$\bar y_{\mu\nu}\equiv \frac{1}{2}\left(y_{\mu\nu}+y_{\nu\mu}-\eta_{\mu\nu}{y_\rho}^\rho\right)$ for any tensor $y_{\mu\nu}$\footnote{Our conventions for raising and lowering indices in the 
second line of (\ref{eq1-1}) are the same as those
in Ref.~\cite{rpf2}.}.
The bare (renormalized) scalar boson mass here is $m_o$($m$) 
and we set presently the small
observed~\cite{cosm1,pdg2008} value of the cosmological constant
to zero so that our quantum graviton, $h_{\mu\nu}$, has zero rest mass.
%We return to the latter point, however, when we discuss phenomenology.
%Here, our normalizations are such that $\kappa=\sqrt{8\pi G_N}$
%where $G_N$ is Newton's constant.
Feynman~\cite{rpf1,rpf2} has essentially worked out the Feynman rules for (\ref{eq1-1}), including the rule for the famous
Feynman-Faddeev-Popov~\cite{rpf1,ffp1a,ffp1b} ghost contribution required 
for unitarity with the fixing of the gauge
(we use the gauge in Ref.~\cite{rpf1},
$\partial^\mu \bar h_{\nu\mu}=0$).
%For more details of this material we refer to Refs.~\cite{rpf1,rpf2}. 
%We turn now directly to the quantum loop corrections
%in the theory in (\ref{eq1-1}).
\par
As we have shown
%\begin{figure}
%\begin{center}
%\epsfig{file=fig4cn.eps,width=140mm}
%\includegraphics[width=80mm]{fig4cn.pdf}
%\end{center}
%\caption{\baselineskip=7mm     Graviton loop contributions to the
%scalar propagator. $q$ is the 4-momentum of the scalar.}
%\label{fig1}
%\end{figure}
in Refs.~\cite{bw1,bw2,bw2a}, the large virtual IR effects
in the respective loop integrals for 
the scalar propagator in quantum general relativity  
can be resummed to the {\em exact} result
%\begin{equation}
%\begin{split} 
$i\Delta'_F(k)=\frac{i}{k^2-m^2-\Sigma_s(k)+i\epsilon}
=  \frac{ie^{B''_g(k)}}{k^2-m^2-\Sigma'_s+i\epsilon}$
%&\equiv i\Delta'_F(k)|_{\text{resummed}}
%&=\frac{i}{k^2-m^2-\Sigma_s(k)+i\epsilon}
%\end{split}
%\end{equation}
for{\small ~~~($\Delta =k^2 - m^2$)} where {\small
%\begin{equation}
%\begin{split} 
%B''_g(k)&= -2i\kappa^2k^4\frac{\int d^4\ell}{16\pi^4}\frac{1}{\ell^2-\lambda^2+i\epsilon}\\
%&\qquad\frac{1}{(\ell^2+2\ell k+\Delta +i\epsilon)^2}\\
$B''_g(k)=\frac{\kappa^2|k^2|}{8\pi^2}\ln\left(\frac{m^2}{m^2+|k^2|}\right)$. }      
%\end{split}
%\label{yfs1} 
%\end{equation}}
The form for $B''_g(k)$ holds for the UV(deep Euclidean) regime\footnote{ By Wick rotation, the identification
$-|k^2|\equiv k^2$ in the deep Euclidean regime gives 
immediate analytic continuation to the result for  $B''_g(k)$
when the usual $-i\epsilon,\; \epsilon\downarrow 0,$ is appended to $m^2$.}, 
so that $\Delta'_F(k)|_{\text{resummed}}$ 
falls faster than any power of $|k^2|$. See Ref.~\cite{bw1} for the analogous result
for m=0. Here, $-i\Sigma_s(k)$ is the 1PI scalar self-energy function
so that $i\Delta'_F(k)$ is the exact scalar propagator. The residual $\Sigma'_s$ starts in ${\cal O}(\kappa^2)$.
We may drop it in calculating one-loop effects. 
When the respective analogs of $i\Delta'_F(k)|_{\text{resummed}}$\footnote{These follow from the spin independence~\cite{sw-sftgrav,bw1,wein-qft} of a particle's coupling to the graviton.} are used for the
elementary particles, all quantum 
gravity loops are UV finite~\cite{bw1,bw2,bw2a}.
\par
Specifically, we extend our resummed propagator results 
to all the particles
in the ST Lagrangian and to the graviton itself
%, working with the
%complete theory
%\begin{equation}
%{\cal L}(x) = \frac{1}{2\kappa^2}\sqrt{-g} \left(R-2\Lambda\right)
%            + \sqrt{-g} L^{\cal G}_{SM}(x)
%\end{equation}
%where $L^{\cal G}_{SM}(x)$ is SM Lagrangian written in diffeomorphism
%invariant form as explained in Refs.~\cite{bw1,bw2a}, 
and show in the Refs.~\cite{bw1,bw2,bw2a} that the denominator for the propagation of transverse-traceless
modes of the graviton becomes ($M_{Pl}$ is the Planck mass)
\begin{equation}
q^2+\Sigma^T(q^2)+i\epsilon\cong q^2-q^4\frac{c_{2,eff}}{360\pi M_{Pl}^2},
\end{equation}
where $c_{2,eff}\cong  2.56\times 10^4$ is defined in Refs.~\cite{bw1,bw2,bw2a}.
We thus get
(we use $G_N$ for $G_N(0)$) 
\begin{equation}
G_N(k)=G_N/(1+\frac{c_{2,eff}k^2}{360\pi M_{Pl}^2}),\;\; g_*=\lim_{k^2\rightarrow \infty}k^2G_N(k^2)=\frac{360\pi}{c_{2,eff}}\cong 0.0442.
\end{equation}
%and compute the UV limit $g_*$ as
%\begin{equation}
%g_*=\lim_{k^2\rightarrow \infty}k^2G_N(k^2)=\frac{360\pi}{c_{2,eff}}\cong 0.0442.\end{equation}
%We stress that this result has no threshold/cut-off effects in it.
%It is a pure property of the known world.
\par
For the dimensionless cosmological constant $\lambda_*$ we isolate~\cite{drkuniv}  $\Lambda$ via the VEV of
Einstein's equation  
%\begin{equation}
$G_{\mu\nu}+\Lambda g_{\mu\nu}=-\kappa^2 T_{\mu\nu}$
%\label{eineq1}
%\end{equation}
in a standard notation. 
%where $G_{\mu\nu}=R_{\mu\nu}-\frac{1}{2}Rg_{\mu\nu}$,
%$R_{\mu\nu}$ is the contracted Riemann tensor, and
%$T_{\mu\nu}$ is the energy-momentum tensor. 
%Writing
%$g_{\mu\nu}=\eta_{\mu\nu}+2\kappa h_{\mu\nu}$
%for $\eta_{\mu\nu}=\text{diag}(1,-1,-1,-1)$, 
%%We compute~\cite{drkuniv} the VEV of (\ref{eineq1})
%%to isolate $\Lambda$. 
We find that a scalar
makes the contribution to $\Lambda$ given by\footnote{We note the
use here in the integrand of $2k_0^2$ rather than the $2(\vec{k}^2+m^2)$ in Ref.~\cite{bw2i}, to be
consistent with $\omega=-1$~\cite{zeld} for the vacuum stress-energy tensor.} $\Lambda_s\cong -8\pi G_N[\frac{1}{G_N^{2}64\rho^2}]$ and that a Dirac fermion contributes $-4$ times $\Lambda_s$ to
$\Lambda$, where $\rho=\ln\frac{2}{\lambda_c}$ with $\lambda_c(j)=\frac{2m_j^2}{\pi M_{Pl}^2}$ for particle j with mass $m_j$.
 The deep UV limit of $\Lambda$ then becomes, allowing $G_N(k)$
to run,
\begin{equation}
%\begin{split}
\Lambda(k) \operatornamewithlimits{\longrightarrow}_{k^2\rightarrow \infty} k^2\lambda_*,\;
\lambda_* =-\frac{c_{2,eff}}{2880}\sum_{j}(-1)^{F_j}n_j/\rho_j^2 \cong 0.0817
%\end{split}
\end{equation} 
where $F_j$ is the fermion number of particle $j$, $n_j$ is the effective
number of degrees of freedom of $j$ and $\rho_j=\rho(\lambda_c(m_j))$.
$\lambda_*$ vanishes in an exactly supersymmetric theory .\par
The UV fixed-point calculated here, 
$(g_*,\lambda_*)\cong (0.0442,0.0817),$ and the estimate
$(g_*,\lambda_*)\approx (0.27,0.36)$
in Refs.~\cite{reuter1,reuter2} are similar in that in both of them
$g_*$ and $\lambda_*$ are 
positive and are less than 1 in size. 
%See Refs.~\cite{bw1} for more discussion of this comparison. 
See Refs.~\cite{bw1} for
further discussion of the relationship between the two fixed-point predictions.
%our $\{g_*,\;\lambda_*\}$ predictions and those in Refs.~\cite{reuter1,reuter2}. 
\par
\section{\bf Review of Planck Scale Cosmology and an Estimate of $\Lambda$}
In the Einstein-Hilbert theory, the authors in Ref.~\cite{reuter1,reuter2}, using the exact renormalization group
for the Wilsonian~\cite{kgw} coarse grained effective 
average action in field space, as discussed in Section 1,  
have argued that the dimensionless Newton and cosmological constants
%the attendant running Newton constant $G_N(k)$ and running 
%cosmological constant
%$\Lambda(k)$ 
approach UV fixed points as the attendant scale $k$ goes to infinity
in the deep Euclidean regime, as we have also found in RQG. 
%Accordingly, we have 
%$k^2G_N(k)\rightarrow g_*,\; \Lambda(k)\rightarrow \lambda_*k^2$
%for $k\rightarrow \infty$ in the Euclidean regime.\par
To make contact with cosmology, one may use a connection between 
the momentum scale $k$ characterizing the coarseness
of the Wilsonian graininess of the average effective action and the
cosmological time $t$.  From this latter connection, the authors
in Refs.~\cite{reuter1,reuter2} arrive at the following extension of the standard cosmological
equations:
%\begin{equation}
\begin{equation}
%\begin{align}
(\frac{\dot{a}}{a})^2+\frac{K}{a^2}=\frac{1}{3}\Lambda+\frac{8\pi}{3}G_N\rho,\;
\dot{\rho}+3(1+\omega)\frac{\dot{a}}{a}\rho=0,\;
\dot{\Lambda}+8\pi\rho\dot{G_N}=0,\;
G_N(t)=G_N(k(t)),\;
\Lambda(t)=\Lambda(k(t)).
\label{coseqn1}
%\end{align}}
\end{equation}
Here, $\rho$ is the density and $a(t)$ is the scale factor
with the Robertson-Walker metric given as
\begin{equation}
ds^2=dt^2-a(t)^2\left(\frac{dr^2}{1-Kr^2}+r^2(d\theta^2+\sin^2\theta d\phi^2)\right)
\label{metric1}
\end{equation}
where $K=0,1,-1$ correspond respectively to flat, spherical and
pseudo-spherical 3-spaces for constant time t.  
The equation of state is 
$ 
p(t)=\omega \rho(t),
$
where $p$ is the pressure.
The relationship between $k$ and the cosmological time $t$ is 
$
k(t)=\frac{\xi}{t}
$
for a constant $\xi>0$ determined
from constraints on
physical observables.\par 
The authors in Refs.~\cite{reuter1,reuter2}, using the UV fixed points for $k^2G_N(k)\equiv g_*$ and
$\Lambda(k)/k^2\equiv \lambda_*$ obtained independently, solve the cosmological system given above. For $K=0$, they find
a solution in the Planck regime where $0\le t\le t_{\text{class}}$, with
$t_{\text{class}}$ a ``few'' times the Planck time $t_{Pl}$, which joins
smoothly onto a solution in the classical regime, $t>t_{\text{class}}$,
which coincides with standard Friedmann-Robertson-Walker phenomenology
but with the horizon, flatness, scale free Harrison-Zeldovich spectrum,
and entropy problems all solved purely by Planck scale quantum physics.
We now review how to use the Planck scale cosmology of Refs.~\cite{reuter1,reuter2} and the UV limits $ \{g_*,\; \lambda_*\}$ in RQG~\cite{bw1,bw2,bw2a} in Refs.~\cite{bw2i} to predict~\cite{drkuniv} the current value of $\Lambda$.
\par
Specifically, the Planck scale cosmology description of inflation in Ref.~\cite{reuter2}  gives the transition time between the Planck regime and the classical Friedmann-Robertson-Walker(FRW) regime as $t_{tr}\sim 25 t_{Pl}$. 
%(We discuss in Ref.~\cite{drkuniv} the uncertainty of this choice of $t_{tr}$ and we present more on this uncertainty below.)
Starting with the quantity $\rho_\Lambda(t_{tr}) \equiv\frac{\Lambda(t_{tr})}{8\pi G_N(t_{tr})}$ we show in Ref.~\cite{drkuniv} that we get, 
%\begin{equation}
%\begin{split}
%\rho_\Lambda(t_{tr}) \equiv\frac{\Lambda(t_{tr})}{8\pi G_N(t_{tr})}
%         =\frac{-M_{Pl}^4(k_{tr})}{64}\sum_j\frac{(-1)^Fn_j}{\rho_j^2}
%\end{split}
%\label{eq-rho-lambda}
%\end{equation}
employing the arguments in Refs.~\cite{branch-zap} ($t_{eq}$ is the time of matter-radiation equality),  
\begin{equation}
\begin{split}
\rho_\Lambda(t_0)&\cong \frac{-M_{Pl}^4(1+c_{2,eff}k_{tr}^2/(360\pi M_{Pl}^2))^2}{64}\sum_j\frac{(-1)^Fn_j}{\rho_j^2}
          \times \frac{t_{tr}^2}{t_{eq}^2} \times (\frac{t_{eq}^{2/3}}{t_0^{2/3}})^3\cr
%         & \qquad\qquad {\Color{Magenta}\Updownarrow} \qquad\qquad\qquad {\Color{Brown}\Updownarrow} \cr
%         &\qquad {\Color{Magenta}\text{Rad. Dom.}} \qquad {\Color{Brown}\text{Mat. Dom.}}\qquad\qquad {\Color{PineGreen}\Rightarrow}\cr
   & \cong \frac{-M_{Pl}^2(1.0362)^2(-9.194\times 10^{-3})}{64}\frac{(25)^2}{t_0^2}
   \cong (2.4\times 10^{-3}eV)^4.\cr
\end{split}
\label{eq-rho-expt}
\end{equation}
$t_0\cong 13.7\times 10^9$ yrs  is the age of the universe. 
%In the estimate in (\ref{eq-rho-expt}), the first factor in the second line comes from the radiation dominated period from
%$t_{tr}$ to $t_{eq}$ and the second factor
%comes from the matter dominated period from $t_{eq}$ to $t_0$ 
%\footnote{The method of the operator field forces the vacuum energies to follow the same scaling as the non-vacuum excitations.}.
The estimate in (\ref{eq-rho-expt}) is close to the experimental result~\cite{pdg2008}\footnote{The analysis in Ref.~\cite{sola2} also gives 
a value for $\rho_\Lambda(t_0)$ that is qualitatively similar to this experimental result.} 
$\rho_\Lambda(t_0)|_{\text{expt}}\cong ((2.37\pm 0.05)\times 10^{-3}eV)^4$. 
\par
%We do believe our estimate 
%of $\rho_\Lambda(t_0)$
%represents some amount of progress in
%the long effort to understand its observed value  
%in relativistic quantum field theory. We do not consider the estimate to be a precision prediction,
%as hitherto unseen degrees of freedom, such as a high scale GUT theory, 
%may exist that have not been included in the calculation.\par
The three issues of the effect of various spontaneous symmetry breaking energies on $\Lambda$, the effect of our approach to $\Lambda$ on big bang nucleosynthesis(BBN)~\cite{bbn}, and the effect of the time dependence of $\Lambda$ and $G_N$ on the covariance~\cite{bianref1,bianref2,bianref3} of the theory are discussed in detail in Ref.~\cite{drkuniv}. We refer the reader to the discussions in Ref.~\cite{drkuniv}, respectively. Concerning the issue of the error on our estimate, we have argued in Ref.~\cite{eh-consist} that the Heisenberg uncertainty principle, taken together with the structure of the solutions of Einstein's 
% (\ref{eineq1})
equation, implies the constraint $
%\begin{equation}
k\ge \frac{\sqrt{5}}{2w_0}=\frac{\sqrt{5}}{2}\frac{1}{\sqrt{3/\Lambda(k)}}
%\label{eheq3}
%\end{equation}
$
where $\Lambda(k)$  follows from (\ref{eq-rho-expt}) (see  Eq.(52) in Ref.~\cite{drkuniv}). This constraint's equality gives the estimate~\cite{drkuniv}
of the transition time, $t_{\text tr}=\alpha/M_{Pl}=1/k_{\text{tr}}$, from the Planck scale inflationary regime~\cite{reuter1,reuter2} to the Friedmann-Robertson-Walker regime via the implied value of $\alpha$. When we solve for $\alpha$ we get $
%\begin{equation}
\alpha\cong 25.3,$
%\label{eheq5}
%\end{equation}
which agrees with the value $\alpha\cong 25$ implied by the numerical studies in Ref.~\cite{reuter1,reuter2}. This agreement implies an error on $t_{\text tr}$ at the level of a factor ${\cal O}(3)$ or less and an uncertainty on $\Lambda$ reduced from a factor of ${\cal O}(100)$~\cite{drkuniv} to a factor of ${\cal O}(10)$.\par
One may ask what would happen to our estimate if there were a susy GUT theory at high scales?  For definiteness and purposes of illustration, in Ref.~\cite{drkuniv}
we use the susy SO(10) GUT model in Ref.~\cite{ravi-1}
to illustrate how such a theory might affect our estimate of $\Lambda$. We show that one needs a very high mass for the gravitino or that one needs twice the usual particle content with the susy partners of the new quarks and leptons at masses much lower than their partners' masses -- see Ref.~\cite{drkuniv}. We thank Profs. S. Bethke and W. Hollik for the support and kind hospitality of the Werner-Heisenberg-Institut, MPI, Munich, Germany.\par

%\section*{Note Added:}
%Here, we point out for clarity that in computing $\Lambda$
%in the Planck regime the assumption of $K=0$ is presumed 
%as that is the only case for which the Bonanno-Reuter Planck scale
%cosmology has been shown to allow a smooth connection from
%the Planck regime for times near or earlier than the Planck time
%to the semi-classical FRW regime for times after $t_{tr}$.
%For $K=0$, by definition, equal time slices are flat 3-spaces, exactly
%as we have employed in the vacuum states used to compute 
%the zero-point energies that comprise $\Lambda$. Thus the results
%in Sections 3 and 4 are fully self-consistent.

%\section{...}

%\begin{thebibliography}{99}
%\bibitem{...}
%....

%\end{thebibliography}

\end{document}